\documentclass[superscriptaddress,twocolumn,showpacs,preprintnumbers]{revtex4}

\usepackage{graphicx}
\usepackage{dcolumn}
\usepackage{bm}

\begin{document}

\preprint{FERMILAB-Pub-02/015-E}

\title{A High Statistics Measurement of the $\Lambda_c^+$ Lifetime}

\author{J.~M.~Link}
\author{M.~Reyes}
\author{P.~M.~Yager}
\affiliation{University of California, Davis, CA 95616}
\author{J.~C.~Anjos}
\author{I.~Bediaga}
\author{C.~G\"obel}
\author{J.~Magnin}
\author{A.~Massafferi}
\author{J.~M.~de~Miranda}
\author{I.~M.~Pepe}
\author{A.~C.~dos~Reis}
\affiliation{Centro Brasileiro de Pesquisas F\'isicas, Rio de Janeiro,
RJ, Brazil}
% ...


\collaboration{FOCUS Collaboration}
\noaffiliation

\date{\today}

\begin{abstract}
A high statistics measurement of the $\Lambda_c^+$ lifetime from
the Fermilab fixed-target FOCUS photoproduction experiment is
presented.
We describe the analysis technique with particular attention to the
determination of the systematic uncertainty.
The measured value of 
$204.6\pm 3.4~(\mathrm{stat.}) \pm 2.5~(\mathrm{syst.})$~fs from 
$8034\pm122$ $\Lambda_c\rightarrow pK\pi$ decays
represents a significant improvement over the present world average.
\end{abstract}

\maketitle

Experimental measurements of charm particle lifetimes have been used
in the study of strong interaction physics. The
measurements provide some guidance for theoretical calculations of 
non-perturbative strong interaction processes. 
The steady improvement in the precision of the measurements has
not only helped to improve our theoretical understanding of strong 
interactions, but also to help stimulate the development of better theoretical 
tools. These have progressed from the spectator model to
various quarks models and currently to Heavy Quark Expansion 
methods \cite{Reference:bellinibigi}.
These calculational tools are the same or similar to those used in
other areas, for example to determine
the size of the $V_{ub}$ CKM element through inclusive
semileptonic $B$ decays \cite{Reference:incsemilepb}. 
More precise measurements of all of the charm
particle lifetimes will help continue this process of improvement
and extension of applicability.

Precise charm lifetime measurements are now beginning to emerge from
$e^+e^-$ collider experiments 
\cite{Reference:cleodlt,Reference:cleolclt}.
The effects of lifetime and
vertex resolution are also important in mixing and CP violation
measurements \cite{Reference:cleomix1,Reference:belleycp}. 
It is crucial to have accurate lifetime measurements
from fixed-target experiments to act as a standard to
evaluate any relative systematic differences.
The $\Lambda_c^+$ lifetime presented in this paper represents the
most accurate measurement of this quantity to date and is a significant
improvement over the present world average.

The data used were collected by the FOCUS collaboration
in the 1997 fixed-target run at Fermi National Accelerator
Laboratory. The FOCUS spectrometer is an upgrade of the spectrometer used
in the E687 photoproduction experiment \cite{Reference:e687nim1}.
The vertex region consists of four BeO targets and 16 planes of
silicon strip detectors (SSD). 
Two of the SSD planes were placed immediately downstream of
the second target, and two immediately downstream of the fourth
(most downstream) target.
Momentum analysis was made possible by the use of 5
multiwire proportional chambers
and two magnets with opposite polarities. Hadronic particle identification
was achieved using three multicell threshold \v Cerenkov counters
\cite{Reference:focuscerenkovnim}. The data for this measurement were
taken using a photon beam with average energy of $\sim 180$~GeV for
triggered events.

The $\Lambda_c^+\rightarrow pK^-\pi^+$\footnote{The charge conjugate
mode is implicitly implied unless otherwise stated.}\ candidates are
reconstructed using a candidate driven algorithm which is 
highly efficient for all decays including short
lived ones. All $pK^-\pi^+$ candidates are tested to see if they form a
vertex with a confidence level greater than 1\%. The candidate
$\Lambda_c^+$ momentum vector is then projected to search for a production
vertex with one or more tracks. As many tracks as possible
are included in the production vertex
so long as the vertex confidence level is larger than 1\%.
The production
vertex is required to be within one of the four targets.
The separation $L$ between the production and decay vertices is
required to be larger than $6\sigma_L$ where $\sigma_L$ is the
error on $L$ calculated on a candidate-by-candidate basis.
In addition,
each track in the $pK^-\pi^+$ candidate combination must also
satisfy the appropriate \v Cerenkov particle identification criteria.

\begin{figure}
\includegraphics[height=1.5in]{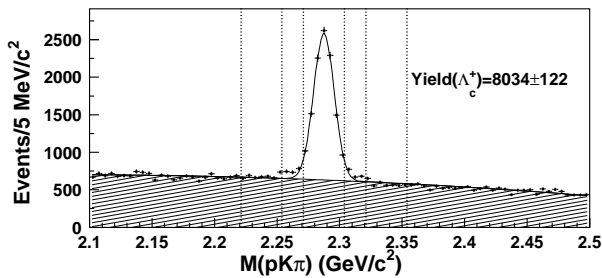}
\caption{\label{fg_pkpimass1} $pK\pi$ invariant mass plot for data (points)
fitted with a Gaussian signal and quadratic background (solid line). 
The shaded area indicates the fitted level of background.
The vertical dotted lines indicate
the signal and sideband regions (see text) used in the lifetime analysis.}
\end{figure}

The $pK\pi$ invariant mass plot for data is shown in Fig.~\ref{fg_pkpimass1}.
The fit shown uses a Gaussian signal and a quadratic background function
which yields $8034\pm 122$ reconstructed 
$\Lambda_c$ decays. The lifetime analysis uses
$pK\pi$ candidates within the signal and symmetric sideband regions as
shown in the figure. All three regions are $4\sigma_m$ wide and the 
centers of the sideband
regions are located $\pm 6\sigma_m$ from the mean of the fitted Gaussian,
where $\sigma_m=8.2$~MeV/$c^2$ is the width of the fitted Gaussian.

For the lifetime analysis we use the reduced proper time,
$t^{\prime}=(L-6\sigma_L)/\beta\gamma c$ \footnote{The reduced proper
time has the clock started at the minimum allowed time for each decay
candidate. In the absence of other corrections, it does not matter where
along the decay exponential one starts the clock, hence the reduced
proper time also follows a pure exponential.},
where
$\beta\gamma=p_{\Lambda_c}/m_{\Lambda_c}$ and require it to be less than
1~ps to reduce long-lived backgrounds. This requirement was already
made for the data shown 
in Fig.~\ref{fg_pkpimass1}. The use of the reduced proper time
ensures that only a small acceptance correction to the
lifetime distribution is needed. The average proper time resolution 
for this decay sample (42~fs)
is small enough compared to the lifetime to use a binned likelihood
method \cite{Reference:e687lclt}.

The $t^{\prime}$ distributions for the decays in the signal and sideband
regions are binned into two separate histograms from 0--1~ps in 20~fs
bins. The observed number of decays in the 
{\em i}$^\mathrm{th}$ $t^{\prime}$ bin
is $s_i$ for the signal region and $b_i$ for the sideband region. The
$t^{\prime}$ distribution of the sideband region is used as a measure of
the lifetime distribution of background events in the signal region. Thus
the expected number of decays in the {\em i}$^\mathrm{th}$ 
$t^{\prime}$ bin of the
signal region is given by:

\begin{eqnarray}
\begin{array}{c}{\rm Expected}\\ {\rm Events}\end{array}=n_i=
S{f(t_i^{\prime})e^{-t_i^{\prime}/{\tau}}\over
\sum_i f(t_i^{\prime})e^{-t_i^{\prime}/{\tau}}}+
{B}{b_i\over\sum_i b_i}.
\label{eq:expectednumber}
\end{eqnarray}
The likelihood that is maximized in the fit is given by
\begin{eqnarray}
{\rm Likelihood}=\prod_i{n_i^{s_i}e^{-n_i}\over s_i!}\times
{(\alpha B)^{N_b}e^{-\alpha B}\over N_b!}
\label{eq:likelihood}
\end{eqnarray}
where $S$ is the total number of signal events and $B$ is the total number of
background events in the signal region and
$S+B=\Sigma s_i$. The total number of events in the sideband region is
$N_b=\Sigma_i b_i$ and $\alpha$ is the ratio of the number of events in the
sideband region to the number of background events in the signal region.
The value of $\alpha$ is obtained from the fit to the invariant mass 
distribution and is very close to $2$. $B$ and $\tau$ are
the fit parameters.

\begin{figure}
\includegraphics[height=1.5in]{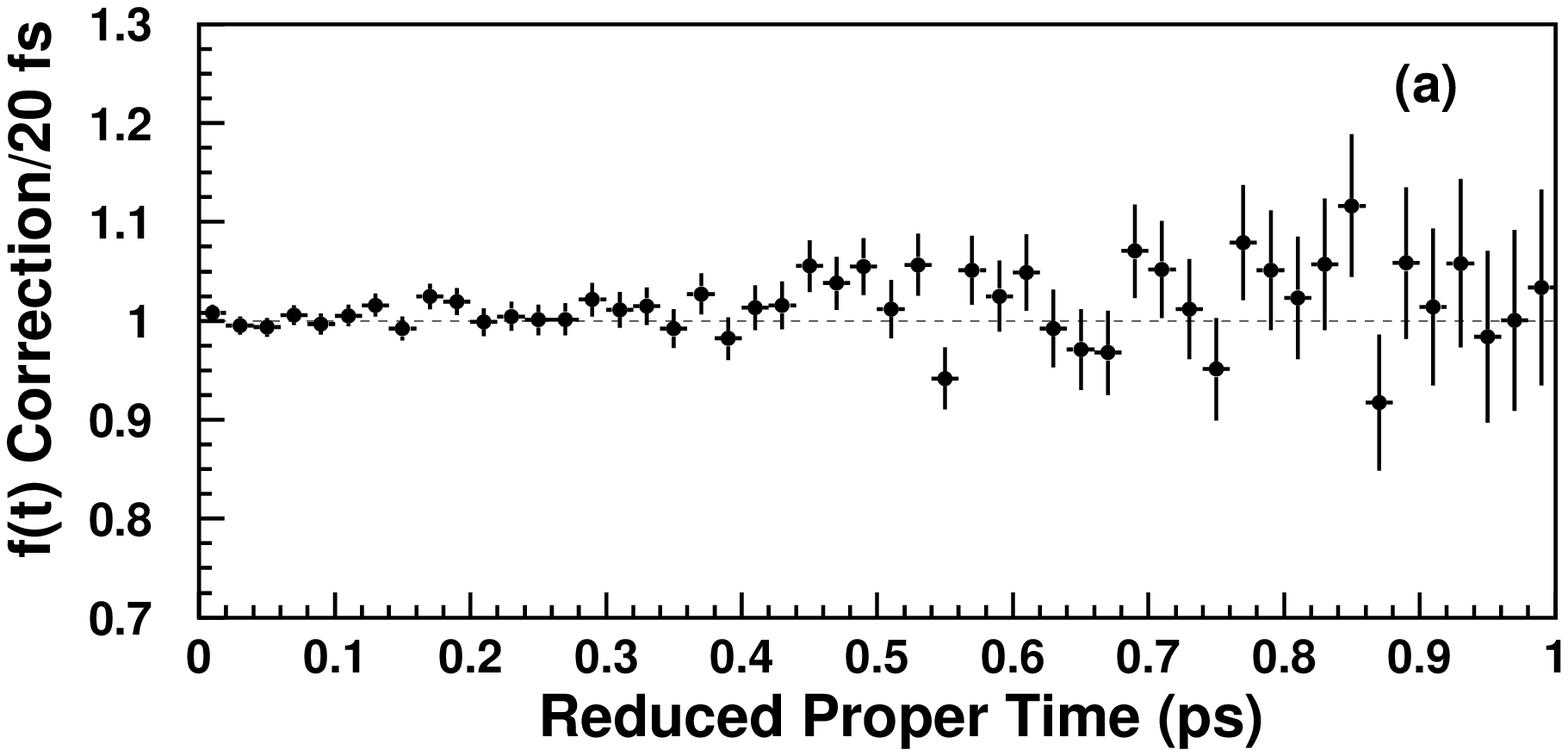}\\
\includegraphics[height=1.5in]{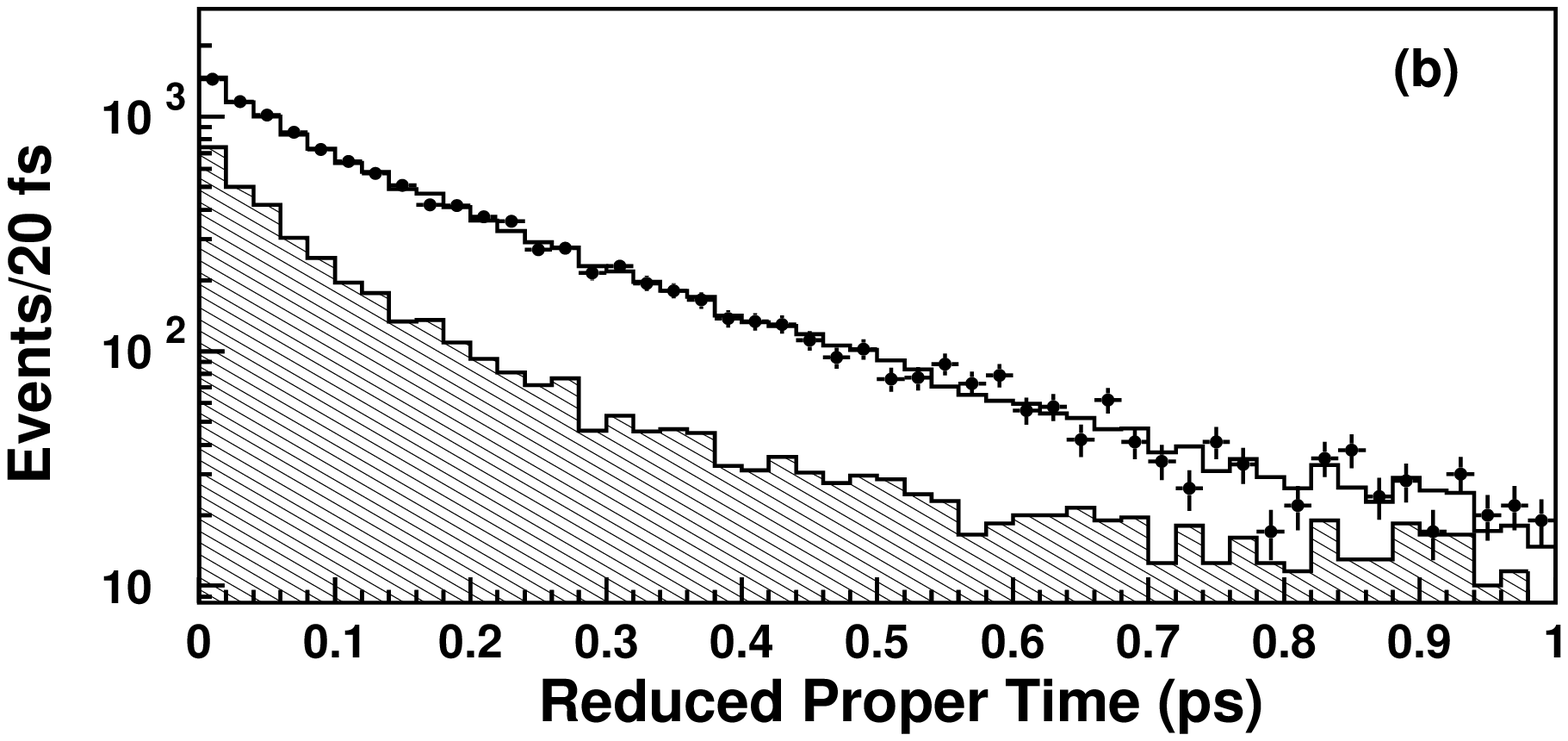}\\
\includegraphics[height=1.5in]{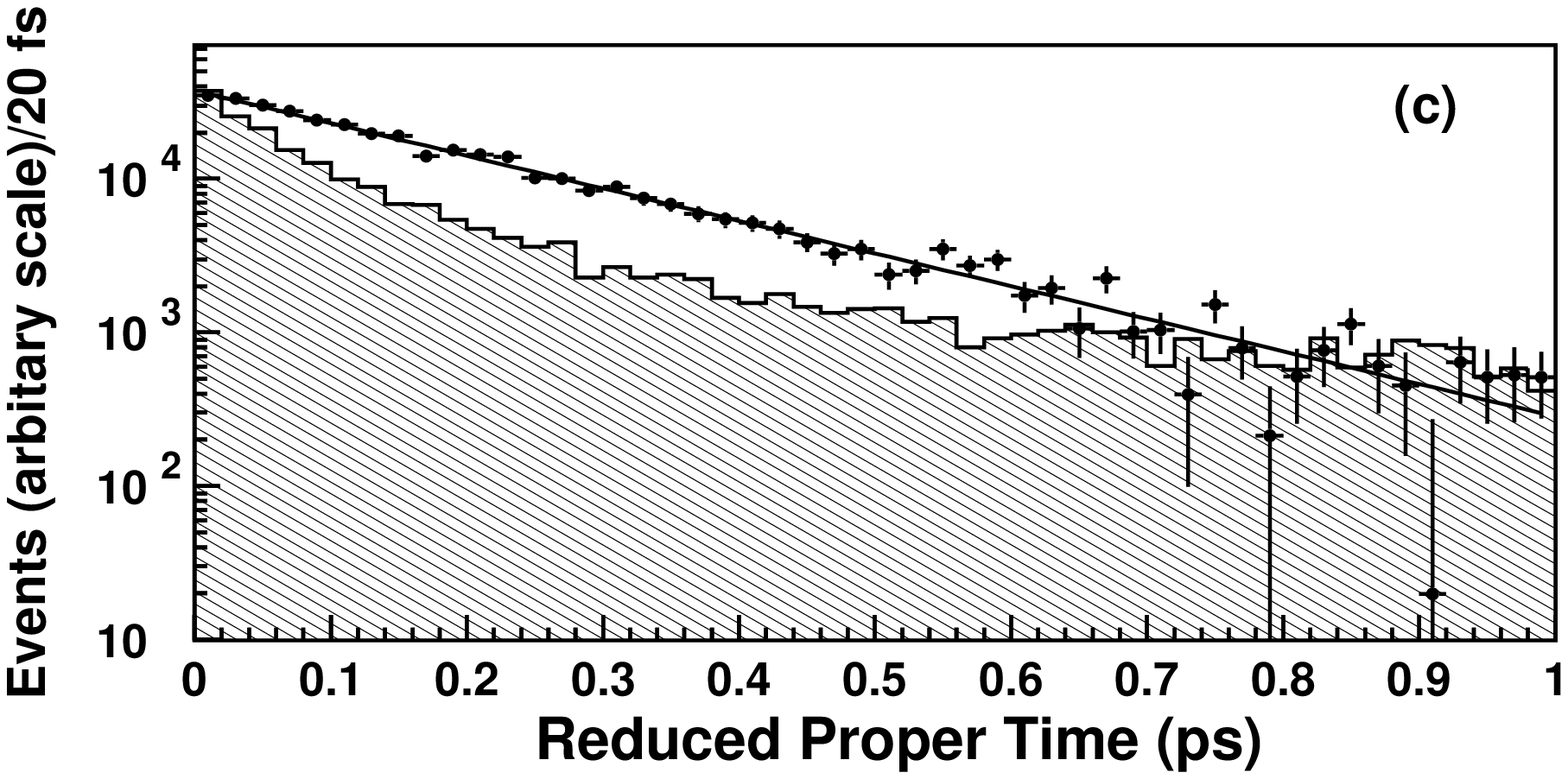}
\caption{\label{fg_ftandfit} (a) The $f(t^{\prime})$ correction function.
Deviation from a flat line
indicates the correction from a pure exponential; (b) The 
lifetime distribution for all decays in the data signal region (points), and
the fit (histogram). The shaded distribution shows the
lifetime distribution of the background component in the signal region;
(c) The lifetime distribution for $\Lambda_c$ decays (points),
{\em i.e.} the sideband subtracted and $f(t^{\prime})$ corrected yield. 
The line is a pure exponential with the fitted lifetime and the shaded region
gives the background.
An arbitary yield scale is used because of the 
particular normalization of $f(t^{\prime})$.}
\end{figure}

The effects of geometrical acceptance, 
detector and reconstruction efficiencies,
and absorption are given by the $f(t^{\prime})$
correction function. The $f(t^{\prime})$ is determined using a detailed
Monte Carlo (MC) simulation of the experiment where the production
(using \textsc{Pythia}~\cite{Reference:pythia61})
was tuned so that the production distributions for data and MC matched.
Note that only the shape of the $f(t^{\prime})$ function is
important and it is obtained by dividing the observed MC $t^{\prime}$
distribution by a pure exponential with the MC generated lifetime.
The $f(t^{\prime})$ distribution is shown in Fig.~\ref{fg_ftandfit}(a).

Using the likelihood function
given above
we obtained a fitted lifetime of $204.6\pm 3.4$~fs. The lifetime
distribution of all decays in the signal region is shown in
Fig.~\ref{fg_ftandfit}(b)\ together with the fit and the level of
background contained in the signal region.

Detailed studies were performed to determine the systematic uncertainty in 
this measurement.

The uncertainty in the absolute time scale was investigated by studying
the absolute length and momentum scales in the experiment. For the
length scale, comparisons were made between
measurements of the distances
between silicon planes in the target region.
The values obtained using vertex positions in the data with the
standard vertexing code agree well with those obtained using
precision instruments.
The absolute momentum and mass
scales were checked by comparing the reconstructed masses of
charm and strange 
mesons and hyperons with established values. 
Our studies showed no evidence of any scale offset, but due to the limited
statistical precision of these comparisons
we assign an uncertainty of $\pm 0.11$\%\ to the absolute time scale.


The backgrounds are composed of a non-charm and a charm component;
these two background components
are approximately equal in our sample and
fairly evenly distributed across the signal and sideband mass regions.
The level and lifetime distribution of the background in the signal
mass region is assumed to be well represented by symmetric mass sidebands 
close to the signal region. The uncertainties that 
arise because of these assumptions were determined by a large number of
studies.

The contamination from
$D^+\rightarrow K^-\pi^+\pi^+$,
$D^+\rightarrow K^-K^+\pi^+$ and $D_s^+\rightarrow K^-K^+\pi^+$ decays
misidentified as $pK^-\pi^+$ decays were determined in our sample.
We loosened the \v Cerenkov requirements on the data and used the MC 
efficiencies to extrapolate to tighter particle identification criteria. 
From this we found the above three decays respectively contribute
0.5\%, 1.3\%\ and 2.7\%\ of
the total background in the signal region.
The small contribution of
these reflection backgrounds and the fact that they are distributed
fairly uniformly across the signal and sideband mass regions mean they
give rise to insignificant uncertainties. This was verified in a test by
explicitly eliminating them by cutting out the appropriate mass regions.
Using variations in particle identification and vertexing selection to 
significantly change the signal/background ratio also showed no
significant uncertainties.

The background lifetime uncertainty was further investigated by using
symmetric sidebands of different widths ($4$--$16\sigma_m$), and located
at different separations from the signal region ($\pm 4$ to
$\pm 16\sigma_m$). The effect of using only the low or only the high mass
sideband was also studied. 
The effect of having the fit parameter $B$ truly free by
eliminating the background term in the likelihood 
(second term in Eq.~(\ref{eq:likelihood})) was studied and
found to be inconsequential. Note that the results of the $pK\pi$ mass fit are
only used in the background term in the likelihood.

Finally, an independent analysis which did not rely on
knowledge of the background lifetime distribution was performed. In this
analysis the data were split into twenty 50~fs wide reduced proper time 
bins from 0--1~ps. The number of $\Lambda_c^+\rightarrow pK^-\pi^+$
decays in each bin was determined in a mass fit and the yields fitted
to an exponential decay distribution modified by a
$f(t^{\prime})$ correction function. This $f(t^{\prime})$ function was
obtained separately for this analysis from the MC, doing the same
split into twenty time bins and fitting the mass distributions for each 
MC bin.
This $f(t^{\prime})$ correction function agrees well with that
obtained in the standard analysis method.

From these studies we assign a background systematic uncertainty of
$\pm 0.77$\%.

Uncertainties in the $f(t^{\prime})$ correction include uncertainties from
the geometrical acceptance, the detector and reconstruction
efficiencies, the production model,
the absorption cross-sections, and the decay dynamics.

With our chosen selection criteria, the $f(t^{\prime})$ correction
reduces the fitted lifetime by 1.19\%. A number of studies were performed
to study the uncertainty in this correction. Since the correction function
is obtained from MC simulations, care was taken to ensure that this
simulation correctly reproduces a very large number of data distributions.
In particular the MC reproduces the data $\Lambda_c^+$ longitudinal and
transverse momenta, the multiplicity of the production vertex, and the decay
length and proper time resolutions. A sensitive check of the 
acceptance and efficiency part of the MC
correction was done using high statistics $K_S^0\rightarrow\pi^+\pi^-$ decays.
Short-lived $K_S^0$ decays were reconstructed using the same analysis
methods in the same decay region as the $\Lambda_c^+$ decays. Since the
$K_S^0$ lifetime is well known we can determine the $f(t^{\prime})$ correction
in data and compare it to that obtained in our MC simulation. The agreement
is excellent but was limited by both data and MC statistics to a sensitivity
of $\pm 2$\%\ of the correction. 
Using this as the level of the uncertainty in the
$f(t^{\prime})$ correction, we can assign a systematic uncertainty due to
this correction of $\pm 0.83$\%.
Possible time dependent systematic effects were looked for by splitting the
data into different time periods and comparing the fitted lifetimes.
We also compared the
separate fitted
lifetimes for decays originating from each of the four targets.
No systematic uncertainties were found in these two comparisons.

Our limited knowledge of the production and decay of the
$\Lambda_c^+$ could contribute to a systematic uncertainty.
This was studied using
different MC simulations where the production parameters and the
resonance substructure of the decay were varied over reasonable ranges.
Production systematics were also studied by splitting the data into
different bins of longitudinal and transverse $\Lambda_c^+$ momenta,
primary vertex multiplicity, and by comparing the fitted lifetimes for
particles and anti-particles. We assign a systematic uncertainty of
$\pm 0.38$\%\ due to our limited knowledge of $\Lambda_c^+$ production
and decay.

In order to use the reduced proper time we must be able to correctly
model our proper time resolution.
This was verified by comparing the distributions
for data and MC and by studying splits of the data sample 
that can be sensitive
to resolution effects. The data were split into bins of proper time
resolution and reconstructed invariant mass. Variations of the proper time
bin width
from 10 to 100~fs were also studied as was changing the fitted range
from 0--0.6~ps to 0--1.4~ps, and from
0--1~ps to 0.2--1~ps. We assign a systematic uncertainty of
$\pm 0.12$\%\ to the lifetime due to resolution uncertainties.

The systematic uncertainty due to absorption of the 
$\Lambda_c^+$ and daughter particles was studied by varying the
charm interaction cross-section by 100\%\ and the
daughter particle interaction cross-sections by 50\%\ 
in the MC. It was
also studied by comparing the lifetimes of decays occuring inside
and outside of the target, and by comparing the lifetimes for decays
where the $\Lambda_c^+$ was produced in the upstream half of each target
with those produced in the downstream half of the same target.
We determined a systematic uncertainty of $\pm 0.23$\%\ due to absorption.

Contributions to the systematic uncertainty are summarized in 
Table \ref{tb_syst}. Taking contributions to be uncorrelated we
obtain a total systematic uncertainty of $\pm 1.23$\%\ or $\pm 2.5$~fs.

\begin{table}
\caption{\label{tb_syst}Contributions to the systematic uncertainty.}
\begin{ruledtabular}
\begin{tabular}{lc}
Contribution & Systematic (\%) \\
\hline
Time scale  & $\pm 0.11$ \\
Backgrounds & $\pm 0.77$ \\
Acceptance  & $\pm 0.83$ \\
Production  & $\pm 0.38$ \\
Resolutions & $\pm 0.12$ \\
Absorption  & $\pm 0.23$ \\
\hline
Total       & $\pm 1.23$ \\
\end{tabular}
\end{ruledtabular}
\end{table}

\begin{table}
\caption{\label{tb_compare}Comparison of recent 
$\Lambda_c^+$ lifetime measurements.}
\begin{ruledtabular}
\begin{tabular}{lcc}
Experiment & Type & $\tau(\Lambda_c^+)$~fs \\
\hline
E687 \cite{Reference:e687lclt}     & FT       & $215\pm 16\pm 8$ \\
SELEX \cite{Reference:selexlclt}   & FT       & $198.1\pm 7.0\pm 5.6$ \\
CLEO II.5 \cite{Reference:cleolclt}& $e^+e^-$ & $179.6\pm 6.9\pm 4.4$ \\
FOCUS (this result)                & FT       & $204.6\pm 3.4\pm 2.5$ \\
\end{tabular}
\end{ruledtabular}
\end{table}

We have measured the $\Lambda_c^+$ lifetime to be
$204.6\pm 3.4~(\mathrm{stat.}) \pm 2.5~(\mathrm{syst.})$~fs using 
$8034\pm122$ $\Lambda_c\rightarrow pK\pi$ decays from the Fermilab FOCUS
photoproduction experiment. This measurement represents a significant
improvement in accuracy and special care was taken to investigate
and properly quantify possible systematic uncertainties.
Table~\ref{tb_compare}\ compares our measurement with previous recent
published results. The difference between this measurement and the
measurement from the CLEO $e^+e^-$ experiment may point to 
the emergence of possible relative systematic 
effects \cite{Reference:hf9cheung}. 
Any such systematic
difference would be important to resolve given the number of recent and future
mixing and CP-violation measurements
that rely on accurate knowledge of lifetime distributions.

We wish to acknowledge the assistance of the staffs of Fermi National
Accelerator Laboratory, the INFN of Italy, and the physics departments of the
collaborating institutions. This research was supported in part by the U.~S.
National Science Foundation, the U.~S. Department of Energy, the Italian
Istituto Nazionale di Fisica Nucleare and Ministero dell'Universit\`a e della
Ricerca Scientifica e Tecnologica, the Brazilian Conselho Nacional de
Desenvolvimento Cient\'{\i}fico e Tecnol\'ogico, CONACyT-M\'exico, the Korean
Ministry of Education, and the Korean Science and Engineering Foundation.

\end{document}